\documentclass[nonacm, sigconf]{acmart}
\AtBeginDocument{%
  }

\setcopyright{none}
\copyrightyear{2025}
\acmYear{2025}
\acmDOI{XXXXXXX.XXXXXXX} 
\acmConference[]{Augmenting Collaborative Problem-Solving: Exploring the Design and Use of GenAI for Groupwork Workshop}{Woodstock, NY}
  
\acmISBN{978-1-4503-XXXX-X/2018/06}

\AtBeginDocument{%
  \fancypagestyle{firstpagestyle}{%
    \fancyhf{}
    \fancyhead[L]{%
      \parbox{\dimexpr\textwidth-2em\relax}{\footnotesize
      Workshop paper presented at the Augmenting Collaborative Problem-Solving: Exploring the Design and Use of GenAI for Groupwork Workshop, held in conjunction with the 28th ACM SIGCHI Conference on Computer-Supported Cooperative Work \& Social Computing Conference in Bergen, Norway, on October 18, 2025.
      }%
    }%
    \fancyhead[R]{}%
  }
}

\begin{document}

\title{Fostering Epistemic Vigilance Towards LLMs in Group Settings}
\title{Confirmation Bias as a Cognitive Resource in LLM-Supported Deliberation}

\author{Sander de Jong}
\orcid{0000-0002-2591-3805}
\affiliation{%
  \institution{Aalborg University}
  \city{Aalborg}
  \country{Denmark}
}
\email{sanderdj@cs.aau.dk}

\author{Rune Møberg Jacobsen}
\orcid{0000-0002-1877-1845}
\affiliation{%
  \institution{Aalborg University}
  \city{Aalborg}
  \country{Denmark}
}
\email{runemj@cs.aau.dk}

\author{Niels van Berkel}
\orcid{0000-0001-5106-7692}
\affiliation{%
  \institution{Aalborg University}
  \city{Aalborg}
  \country{Denmark}
}
\email{nielsvanberkel@cs.aau.dk}

\renewcommand{\shortauthors}{S. de Jong et al.}

\begin{abstract}
Large language models (LLMs) are increasingly used in group decision-making, but their influence risks fostering conformity and reducing epistemic vigilance. Drawing on the Argumentative Theory of Reasoning, we argue that confirmation bias, often seen as detrimental, can be harnessed as a resource when paired with critical evaluation. We propose a three-step process in which individuals first generate ideas independently, then use LLMs to refine and articulate them, and finally engage with LLMs as epistemic provocateurs to anticipate group critique. This framing positions LLMs as tools for scaffolding disagreement, helping individuals prepare for more productive group discussions.
\end{abstract}

\ccsdesc[500]{Human-centered computing~HCI design and evaluation methods}
\ccsdesc[500]{Human-centered computing~Collaborative and social computing design and evaluation methods}

\keywords{Large Language Models, Confirmation Bias, Epistemic Vigilance, Group Decision-Making, Human–AI interaction}



\maketitle
\thispagestyle{firstpagestyle}

\section{Introduction}
As Large Language Models (LLMs) are increasingly integrated into collective decision-making tasks, it is crucial to understand how their presence shapes sensemaking and deliberation processes. Because LLMs often formulate their answers confidently and persuasively, even when incorrect~\cite{Jones2024LiesDamnedLies, Salvi2025ConversationalPersuasivenessGPT4}, they can shape people's reasoning processes. Alarmingly, incorrect AI explanations often convince people even more than correct ones~\cite{Danry2025DeceptiveExplanationsLarge}.

These persuasive effects extend beyond factual tasks, influencing people's political opinions~\cite{Hackenburg2024EvaluatingPersuasiveInfluence, Argyle2023LeveragingAIDemocratic, Bai2023ArtificialIntelligenceCan} and even durably reducing belief in conspiracy theories~\cite{Costello2024DurablyReducingConspiracy}. In group settings, these persuasive characteristics position LLMs in an authoritative role in decision-making processes~\cite{deJong2024GroupAwareness}. Similarly, LLMs can influence individuals' tendency to conform to a majority opinion, with the presence of an LLM in the majority group reducing the group size required for people to conform~\cite{deJong2025Conformity}. Such conformity can lead to convergence of thinking known as `groupthink'~\cite{Janis1972VictimsGroupthinkPsychological}. LLM's tendency to show sycophancy towards people's opinions further enhances this by reinforcing initial ideas and triggering confirmation bias~\cite{Sharma2025UnderstandingSycophancyLanguage}, the tendency to seek and favour information that supports one’s beliefs.

While sycophancy and `groupthink' highlight the risks of persuasive LLMs, cognitive biases do not always undermine reasoning. In individuals or groups of people holding similar views, arguments are often not critically evaluated, and confirmation bias can therefore lead to poor decision outcomes. However, Mercier et al.\ argue that in contexts where people disagree but share a common interest in the truth, confirmation bias contributes to an efficient division of cognitive labour~\cite{Mercier2011WhyHumansReason}. Rather than assessing all sides, each individual focuses on gathering evidence in support of their own position. When these arguments are subsequently presented to the groups, they activate epistemic vigilance in others, a process that helps detect weak, misleading, or flawed reasoning. Together, confirmation bias and epistemic vigilance can contribute to better group-level reasoning outcomes, as long as individuals are exposed to diverse viewpoints and critically evaluate each other's arguments. 

In this paper, we draw on cognitive science literature, particularly the Argumentative Theory of Reasoning, to examine how confirmation bias and epistemic vigilance shape group reasoning. We subsequently outline research directions for leveraging LLMs to help prepare individuals for group discussions.

\section{Related Work}
Mercier et al.\ claim that individual reasoning evolved not primarily to find objective truth individually, but to argue for one's beliefs in social settings and to critically evaluate others' arguments~\cite{Mercier2011WhyHumansReason}. This \emph{Argumentative Theory of Reasoning} suggests that reasoning is fundamentally dialogical rather than solitary, shaped by the social need to persuade and to resist persuasion. According to the theory, reasoning operates through a two-step process: (1) individuals generate arguments to support their own intuitive beliefs, and (2) these arguments are then exposed to the epistemic vigilance of others, who critically assess and challenge them.

Within this theory, confirmation bias is reinterpreted as an adaptive cognitive strategy that works hand-in-hand with epistemic vigilance. It helps individuals generate coherent, persuasive arguments in favour of their own positions, thereby stimulating meaningful disagreement and debate. The authors claim that confirmation bias is not epistemically harmful as long as it operates in an environment of epistemic vigilance, where others challenge and correct flawed reasoning~\cite{Mercier2011WhyHumansReason, Sperber2010EpistemicVigilance}. Hereby, confirmation bias and epistemic vigilance function as complementary mechanisms that can improve group-level reasoning outcomes.

These insights have direct implications for the design of group interactions with LLMs. On one hand, LLMs may suppress epistemic vigilance when users accept their output without critical scrutiny. On the other hand, they can serve as facilitators of human reasoning by helping participants structure their arguments, prompting reflection, guiding the discussion, and surfacing disagreement. In these facilitative roles, LLMs can help create the dialogical conditions needed for epistemic vigilance to emerge without undermining human ownership of reasoning and belief revision. Conversely, LLMs can be considered as poor epistemic challengers during the dialogue itself. Unlike humans, they lack genuine goals, beliefs, or social stakes, making their arguments feel artificial or low-stakes. Furthermore, their perceived authority can lead users to overrely on their outputs, even when they are wrong. This undermines the process of critical scrutiny that epistemic vigilance depends on. Therefore, LLMs may be most effective as facilitators that help humans engage in argumentative interaction with each other rather than as challengers themselves.
 
We therefore focus on three stages of individual evaluation before group discussion, where LLMs can most effectively support reasoning that later activates epistemic vigilance in others. By contrast, the role of LLMs as facilitators of group discussions has already been explored more extensively, using them to play the role of devil's advocate~\cite{Chiang2024EnhancingAIAssistedGroup}, model diverse opinions as personas~\cite{Shi2024ArgumentativeExperienceReducing, Zhang2024SeeWidelyThinka}, and facilitate equal participation in discussions~\cite{Kim2021ModeratorChatbotDeliberative}.

\section{Integrating Confirmation Bias and Epistemic Vigilance in Reasoning}
We present our idea as a three-step process. First, individuals generate ideas independently to preserve epistemic ownership. Second, the LLM helps refine these ideas through articulation and support. Third, the LLM acts as an epistemic provocateur, preparing individuals for confrontation in group discussion.

\subsection{Independent Idea Generation}
To best leverage confirmation bias as a cognitive resource, the process should begin with ideation without LLM involvement, preserving individuals’ epistemic ownership of their opinions. This ownership motivates people to defend their views in subsequent stages. Introducing LLMs too early risks premature convergence around AI-generated content and can lead to uncritical acceptance of outputs, thereby weakening the dialogical process.

In this stage, confirmation bias plays a constructive role: rather than being a flaw, it encourages individuals to selectively search for and organise reasons that support their initial position. This generates sharper, more one-sided arguments, which provide the material for epistemic vigilance to operate when scrutinised by others. By entering the group exchange with personally generated, well-defended positions, individuals help establish the conditions for a productive clash of perspectives rather than passive consensus.

Moreover, starting without LLMs avoids the risk of AI homogenisation, where early suggestions steer participants towards similar framings because they are exposed to similar LLM output. Such convergence undermines the division of cognitive labour that makes group reasoning effective. We therefore present the initial phase as a crucial investment: individuals arrive at later stages with personally owned ideas and arguments sharpened by their own bias-driven search, making them more motivated and better prepared to engage in the clash of perspectives that fuels epistemic vigilance.

\subsection{Articulation and Strengthening}
Once individuals have generated their own initial ideas, LLMs can support articulation and refinement. Confirmation bias is hereby deliberately harnessed instead of suppressed: individuals remain motivated to defend their positions, while the LLM helps them express those positions in a clearer, more persuasive form. By rephrasing arguments in more coherent language and suggesting ways to structure claims, reasons, and evidence, LLMs can ensure that the ideas are comprehensible and compelling to others in the group.

The role of the LLM here is to amplify the individual’s initial stance rather than reshape it. Through light affirmation (i.e., sycophancy) of the individual’s reasoning before critique, the LLM leverages confirmation bias to make participants more invested in defending their positions. In addition, it can help individuals seek supporting data points, analogies, or examples that add weight to their claims without reframing their initial position. This phase is essential for preparing arguments to withstand scrutiny in later stages: poorly articulated ideas are easily dismissed in group settings, regardless of their merit, while well-expressed arguments invite deeper engagement and more critical evaluation.

\subsection{Challenging Beliefs to Refine Ideas}
LLMs can act as epistemic provocateurs, helping individuals test their positions before entering group discussion. By asking probing questions, presenting common counterarguments, or simulating opposition, the system facilitates a rehearsal where participants can get ahead of likely critique and refine their arguments in advance. This process activates content-oriented epistemic vigilance~\cite{Sperber2010EpistemicVigilance}, prompting individuals to evaluate the coherence, plausibility, and strength of their own reasoning, ensuring that they are better prepared to withstand scrutiny once their ideas are exposed to others.

However, the LLM should not take over the role of critic or judge. If individuals rely too heavily on the LLM’s assessments, they may defer to its authority rather than engage in their own critical evaluation. Such overreliance risks reducing epistemic vigilance rather than fostering it. The LLM’s role in this phase is therefore to provoke and prepare instead of decide, leaving the genuine clash of perspectives to emerge within the group itself.

Taken together, these three phases ensure that individuals enter group discussion with arguments they genuinely own, have articulated clearly, and have already tested against simulated critique. By preserving the driving force of confirmation bias to generate arguments, while gradually activating epistemic vigilance in others to evaluate them, LLMs can support individuals in preparing for richer discussions. The goal is to scaffold human disagreement so that when perspectives clash in the group, the conditions for productive reasoning are already in place.







\begin{acks}
This work is supported by the Carlsberg Foundation, grant CF21-0159.
\end{acks}

\bibliographystyle{ACM-Reference-Format}
\bibliography{references}


\end{document}